# High-speed graphene-silicon-graphene waveguide PDs with high photo-to-dark-current ratio and large linear dynamic range

*Jingshu Guo[†,‡,#], Chaoyue Liu[†,#], Laiwen Yu[†], Hengtai Xiang[†], Yuluan Xiang[†], and Daoxin Dai[\*,†,‡,§]*

[†] State Key Laboratory for Modern Optical Instrumentation, Zhejiang Provincial Key Laboratory for Sensing Technologies, College of Optical Science and Engineering, International Research Center for Advanced Photonics, Zhejiang University, Zijingang Campus, Hangzhou, 310058, China.

[‡] Intelligent Optics & Photonics Research Center, Jiaxing Research Institute, Zhejiang University, Jiaxing 314000, China.

[§] Ningbo Research Institute, Zhejiang University, Ningbo 315100, China.

[#] These authors contributed equally to this work: Jingshu Guo, Chaoyue Liu.

[*]Corresponding author e-mail: dxdai@zju.edu.cn.

**ABSTRACT**: Two-dimensional materials (2DMs) meet the demand of broadband and low-cost photodetection on silicon for many applications. Currently, it is still very challenging to realize excellent silicon-2DM PDs. Here we demonstrate graphene-silicon-graphene waveguide PDs operating at the wavelength-bands of 1.55 μm and 2 μm, showing the potential for large-scale integration. For the fabricated PDs, the measured responsivities are respectively ~0.15 mA/W and ~0.015 mA/W for the wavelengths of 1.55 μm and 1.96 μm. In particular, the PDs exhibit a high bandwidth of ~33 GHz, an ultra-low dark current of tens of pico-amperes, a high normalized photo-to-dark-current ratio (NPDR) of $1.63\times10^6$ W$^{-1}$, as well as a high linear dynamic range of 3 μW-1.86 mW (and beyond) at 1.55 μm. According to the measurement results for the wavelength-bands of 1.55/2.0 μm and the theoretical modeling for the silicon-graphene heterostructure, it is revealed that internal photo-emission and photo-assisted thermionic field emission dominantly contribute to the photoresponse in the graphene-silicon Schottky junctions, which helps the future work to further improve the performance.

## INTRODUCTION

Two-dimensional materials (2DMs)[1] feature the wide-range bandgaps, high mobility, and flexible integration, meeting the demands on wide-spectrum, high-performance, and integration-friendly photodetection on silicon photonic chips.[2, 3] Currently there have been several typical silicon-2DM

PDs (PDs) developed,[4] including metal-2DM-metal PDs, metal-2DM+X-metal PDs, and 2DM-heterostructure PDs. It is still very challenging to realize silicon-2DM PDs with high overall-performances on bandwidth, sensitivity, as well as linearity by using low-cost and reproductive fabrication technologies.[4,5] A popular strategy is using the dual-gate FET configuration with graphene *p-n* homojunction, in which way these devices can operate at zero bias with a bandwidth of >10 GHz based on the photo-thermoelectric (PTE) effect.[6,7] Meanwhile, more efforts are still needed to suppress the thermal noise and enhance the photoresponse.[6,7] The silicon-graphene heterostructure[8] provides another promising option for the near/mid-infrared photodetection, because of its potential advantages of easy fabrication and low dark current.[4] Currently various surface-illuminated[9-12] and waveguide-integrated[13,14] silicon-graphene heterostructure PDs have been demonstrated, while most of them have bandwidths below MHz. As an exception, Li *et al.* demonstrated a waveguide-integrated silicon-graphene p-i-n photodiode with a high RF bandwidth estimated from the measured microwave spectrum of the converted photoelectric signal with a center frequency of 40 GHz.[14] In order to improve the PD performances, in the past years the carrier transport mechanisms in various 2DM-2DM and 2DM/bulk-material heterostructures[15,16] have drawn much attention from academia, including silicon-graphene Schottky diodes.[17,18] For a graphene-semiconductor/insulator interface, if light is absorbed by graphene only, the photoresponse mechanisms can be classified to two types, depending on whether the thermal relaxation process of the photo-induced carriers involves. The *first* type is for the case with the photo-thermionic (PTI) effect. Here the photo-excited carriers may form a Fermi–Dirac distribution with an increased temperature $T_e$ higher than the lattice temperature. The high-energy tail of the distribution of the carriers (called "hot carriers") contributes the photo-thermionic current. For instance, the PTI effect was found in the graphene/WSe$_2$/graphene heterostructure,[19] and the plasmonics-enhanced PTI was observed in the graphene/hBN/graphene heterostructure[20]. The *second* type is similar to the internal photo-emission (IPE) effect in the conventional metal-semiconductor Schottky PDs, in which case the photo-excited carriers may go over or tunnel through the energy barrier and then contribute to the photocurrent directly. For the second type, the mechanisms include the prompt internal photoemission (PIPE) in a graphene/SiC Schottky junction [21], the direct tunneling in a graphene/hBN/graphene tunneling junction,[22] the photoelectric effect in a graphene/2D-semiconductor junction,[23] as well as the hot-electron emission in a graphene/vacuum system.[24] Furthermore, it has been shown that the bias voltage,[21,22,24] the incident optical power [21,23,24] and the photon energy[22,23] usually play important roles in the competition of different photoelectric conversion pathways. In order to further develop high-performance silicon-graphene heterostructure PDs, more efforts are still desired to obtain a clearer physical image for light-induced carrier dynamics and photoresponse mechanisms.

In this work, we propose and demonstrate a graphene-silicon-graphene heterostructure PDs with few-layer graphene (FLG). A thin silicon photonic platform is adopted for enhancing light absorption in graphene. In particular, we introduce a multimode-interference (MMI)-based wavelength-division-multiplexer in photonic integrated circuits, so that the present PD working for both wavelength-bands of 1.55/2 μm can be characterized conveniently. The fabricated graphene-silicon-graphene PD typically has a high bandwidth of ~33 GHz, benefiting from short transit time and small RC time constant. The normalized photo-to-dark-current ratios (NPDRs) are as high as $1.63×10^6$ W$^{-1}$ and $2.66×10^5$ W$^{-1}$ at 1.55 μm and 1.96 μm, respectively. According to the experimental results and the theoretical modeling, it is revealed that the IPE effect and the photo-assisted thermionic field emission (PTFE) dominate the photoresponse in the present graphene-silicon-graphene Schottky junctions, showing the potential pathway of the internal quantum efficiency (IQE) improvement in the future.

**RESULTS**

**Design and simulation.** Figure 1a illustrates the schematic configuration of the present FLG-Si-FLG heterostructure waveguide PD. The FLG is divided into two parts with a gap between them (where the gap width is chosen as $w_{gap}$=300 nm in the present design), covering the silicon photonic waveguide symmetrically and forming an FLG-Si-FLG structure. Meanwhile, the FLG parts connect with the corresponding metal electrodes, where the metal-graphene-metal sandwiched contact-structures are introduced for low contact resistance.[25] In order to enhance the light-graphene interaction, here the thin-silicon ridge waveguide[25] is adopted with a silicon height of $h_{si}$=100 nm, an etching height of $h_{etch}$=50 nm, and a silicon ridge width of $w_{si}$=1300 nm.

Figure 1b shows the present silicon photonic integrated circuit (PIC) consisting of a PD, a pair of ~1.55/2 μm two-channel wavelength-division (de)multiplexers, and the corresponding grating couplers for ~1.55/2 μm. Such a PIC configuration supports convenient experimental measurement for the two wavelength-bands of 1.55/2 μm. For each wavelength-band, light is coupled from the fiber to the fundamental TE mode of the input strip silicon photonic waveguide by using the corresponding grating couplers. Then light is routed to the input port of the PD through the wavelength-division multiplexer.

More details on the calibration and the normalization for the optical power are given in Supplementary Note 1. The mode field is calculated from a finite-element method mode-solver tool (COMSOL). For FLG, the optical conductivity is set by scaling the conductivity of single-layer graphene according to the number of layers.[26] The real part of the optical conductivity determining the optical absorption ability of graphene is sensitive to the graphene chemical potential. As discussed below, the graphene doping in this work is light enough ($|\mu_g|\lesssim$0.2 eV) to guarantee full optical absorption ability, in which case the real part of the optical conductivity per layer is close to the universal optical conductivity of $\sigma_0$=60.8 μS for both 1.96 μm and 1.55 μm.[27] In the simulation, we set |$\mu_g$|=0.2 eV to give an estimation. More details about the simulation modeling were given in our previous work.[25] Figure 1c shows the calculated fundamental TE mode fields at the two wavelengths of 1.55/1.96 μm. Regarding that light is absorbed by graphene only in the present case, the graphene absorbance is given by $\eta(L)$=1−$10^{-0.1\alpha L}$, where $\alpha$ is the mode absorption coefficient in dB/μm, $L$ is the length of the graphene sheet. Fig. 1d shows the calculated absorbance $\eta$ at the wavelengths of 1.55/1.96 μm for the cases with different graphene layer numbers $N$ as the length $L$ varies. For the present thin silicon ridge waveguide, the mode fields (see Fig. 1c) and the graphene optical conductivities are similar for the two wavelengths. As a result, the absorption $\alpha$ at the two wavelengths are close. For example, the one-layer graphene has an absorption loss of ~0.111 and 0.107 dB/μm at 1.55 and 1.96 μm, respectively. As shown in Fig. 1d, the absorption coefficient $\alpha$ is proportional to the layer number of graphene. For the case with one-layer graphene, the graphene absorbance $\eta$ reaches >90% when the length $L$>90 μm. In contrast, when the layer number $N$ increases to 3, one has $\eta$>90 % even with the length as short as 30 μm.

**Experimental results and analyses.** The present FLG-Si-FLG heterostructure waveguide PDs were fabricated by the processes of electron-beam lithography, inductively coupled plasma (ICP) etching, graphene transfer, and metal deposition (see details in **Materials and Methods**). Figure 2a shows the fabricated PIC with a 30-μm-long FLG-Si-FLG waveguide PD. The commercial CVD-grown graphene with a few layers was used. As the Raman peak intensity ratio provides a convenient way for graphene number identification, the Raman scattering measurement was carried and the results are given in Fig. 2b. One can see the G-peak at 1584.71 cm$^{-1}$ and the 2D-peak at 2691.99 cm$^{-1}$, and the intensity ratio of 2D-peak to G-peak is about $I_{2D}/I_G$≈2, suggesting that the graphene layer number is ~4.[28] Besides, we characterized the output optical powers from the PIC covered by the graphene sheets with different lengths, and the measured absorption coefficient is estimated to be ~0.4 dB/μm by the cutoff method, which further confirms the conclusion that the graphene layer number is 4. Note that there is usually a native oxide (SiO$_2$) layer formed during the graphene transfer process at the silicon-graphene interface,[29] and our ellipsometer measurement shows that the oxide layer thickness is ~1 nm. The input optical power $P_{in}$ of the PD is estimated by removing the fiber-chip coupling loss and the excess losses from the other passive devices integrated in the PIC. More details about the calibration are given in **Supplementary Note 1**.

**Photoresponse in the silicon-graphene heterostructure.** In order to investigate the photoresponse at the silicon-graphene heterostructure, we also fabricated an FLG-Si waveguide PD for testing on another chip with similar processes.[30] Figure 3a-b shows the band diagrams for two types of mechanisms in the silicon-graphene Schottky diode, including the PTI effect and the IPE effect. Here the spontaneously formed $SiO_2$ oxide layer is included, *n*-doped silicon was used and the electron behavior is mainly considered. Particularly, we define the following two barrier heights, i.e., the Fermi surface barrier height $\Phi_B$ and the Dirac point barrier height $\Phi_D$, which are respectively the barriers referring to the Fermi surface and the Dirac point of graphene. In practice, since the image force lowering (IFL) should be considered, the barriers $\Phi_B$ and $\Phi_D$ become $\Phi_{B-IFL}$ and $\Phi_{D-IFL}$, respectively.[8, 31] Apparently, one has $\Phi_D=\Phi_B+\mu_g$ and $\Phi_{D-IFL}=\Phi_{B-IFL}+\mu_g$, where $\mu_g$ is the chemical potential of graphene.[8]

For the PTI effect shown in Fig. 3a, the photo-excited carriers form a Fermi-Dirac distribution with an increased temperature $T_e$ due to the carrier-carrier scattering, and the hot carriers with sufficient energy can be emitted over the barrier and then contribute to the photoresponse. In this process, the distribution of hot carriers is determined by the Fermi level of the FLG, the density of state (DOS) of the FLG, and the hot carrier temperature $T_e$ jointly. Therefore, the absorbance-normalized responsivity (i.e., $R/\eta$) of the PTI photoresponse is sensitive to $\Phi_{B-IFL}$, other than the energy of the absorbed photons.[19]

Figure 3b shows the direct photon-excited carrier detection mechanisms. Note that the photo-excited carriers have energy $hv/2$ (i.e., half of the photon energy $hv$) higher than the graphene Dirac point due to the unique band structure of graphene; Thus, the IQE of the silicon-graphene PD is determined by $\Phi_D$ (or $\Phi_{D-IFL}$) instead of $\Phi_B$ (or $\Phi_{B-IFL}$), which is totally different from the traditional metal-semiconductor Schottky PDs. As shown in Fig. 3b, only the carriers excited by photons with an energy $hv > 2\Phi_{D-IFL}$ have the chance to be emitted over the Schottky barrier in the IPE process. The photo-carriers without sufficient kinetic energy (strictly speaking, out-plane kinetic energy) still can contribute to the photoresponse by another pathway of tunneling through the barrier, which is named as photo-assisted thermionic field emission (PTFE) process in this paper.

Figure 3c shows measured I-V curves for the fabricated silicon-graphene PD,[30] which has an FLG-Si Schottky diode in the active region. In particular, a subwavelength grating structure is introduced to electrically connect the silicon ridge waveguide and the metal pad, in which way there is no influence on the light propagation in the active region almost. From Fig. 3c, one sees that the I-V curves show standard rectifier behaviors. The current saturation at a high forward bias is attributed to the series resistance and the barrier height increase caused by the graphene doping. When operating with increased reverse bias, the photocurrent increases quickly to be saturated. Here the dark current $I_{dark}$ is about 0.1 nA.

Figure 3d shows the measured responsivity as the reverse bias increases. It can be seen that the results for the wavelengths of 1.55/1.96 μm are with very different responsivities but similar trends. When operating at 1.55 μm, the responsivity increases from 0.02 mA/W to 0.029 mA/W as the bias $|V_{bias}|$ increases from 0 to 1 V, and then reaches ~0.031 mA/W at 5 V. In contrast, for the 1.96 μm wavelength, the measured responsivity increases from ~0 to 6.44×10$^{-4}$ mA/W as the bias |$V_{bias}$| increases from 0.6 V to 2 V, and then reaches 0.008 mA/W at 5 V reserve bias. It shows that the saturated responsivity at 1.55 μm is about 4 folds of that at 1.96 μm. The calculation shows that the light absorptions in the active region (which is 15-μm-long) are 89.95% and 85.76% for the wavelengths of 1.55 μm and 1.96 μm, respectively. It indicates that the absorbance-normalized responsivity and the IQE are very sensitive to the photon energy. On the other hand, as shown in the inset of Fig. 3d, the photocurrent exhibits a linear dependence on the input optical power $P_{in}$ for both wavelengths.

The strong wavelength-dependence of the IQE suggests that the PTI effect is not likely to be the dominated mechanisms for the photoresponse (as discussed above). One may notice that the IQE is pretty low even considering the existence of the 1-nm-thick SiO$_2$ layer. To further figure out the photo-assisted carrier dynamics, we built the band-diagram modeling of the silicon-graphene Schottky diode with the initial parameters confirmed by the Kelvin probe force microscopy (KPFM) measurement (see Supplementary Note 2). As shown in Fig. S1d, the barriers $\Phi_{D\text{-}IFL}$ and $\Phi_{B\text{-}IFL}$ are respectively 0.357 eV and 0.484 eV considering the IFL effect at $^-$5V. For the wavelengths of 1.55 μm and 2 μm, their half photon energies $hv/2$ are 0.4 eV and 0.31 eV, which are respectively higher and lower than the barrier $\Phi_{D\text{-}IFL}$. As a result, the IPE effect and the PTFE process are the dominant mechanisms in the graphene-silicon-graphene heterostructure PD for the wavelengths of 1.55 μm and 1.96 μm, respectively. One should note that the PTFE effect has quite limited quantum efficiency as a tunneling process when compared to the IPE effect. Nevertheless, these two mechanisms both show strong wavelength sensitivity. At the longer wavelength, less photo-excited carriers are excited over the Schottky barrier due to smaller photon energy for the IPE effect, and longer tunneling length leads to lower quantum efficiency for the PTFE process. As the reserve bias increases to >1.5 V, the barrier $\Phi_{D\text{-}IFL}$ and the depletion width in silicon decrease slowly, and consequently the responsivity grows slowly to be saturated for both wavelengths of 1.55 μm and 1.96 μm, as shown in Fig. 3d.

**The graphene-silicon-graphene PD.** As it is well known, traditional silicon-graphene PDs usually need high-quality metal-silicon ohmic contact, which actually is not easy due to the imperfection in the fabrication. Fortunately, no metal-silicon contact is needed for the present graphene-silicon-graphene PD with a simple and fabrication-friendly heterostructure. Furthermore, the present graphene-silicon-graphene PD enables a short carrier transit distance in silicon, providing the potential for high-speed operation because the carrier transit in graphene is usually very fast. Figures 4a, c, e and 4b, d, f show the measured static results of the fabricated graphene-silicon-graphene PD (Device A) with a 60-μm-long graphene sheet for the wavelengths of 1.55 μm and 1.96 μm, respectively. As it can be seen, the measured I-V curves are basically symmetric for both wavelengths of 1.55 μm (Fig. 4a) and 1.96 μm (Fig. 4b). Here we focus on the forward bias operation for more discussions. Figure 4b shows the measured photocurrent $I_{ph}$ at 1.55 μm as the optical power $P_{in}$ varies from 3.10 μW to 1.86 mW (which is the maximal power coupled into the PD with the present measurement-setup). As shown in Fig. 4c, the photoresponse for this graphene-silicon-graphene PD has a very nice linearity even when operating at different bias voltages of 2, 4, and 6 V. When operating at the wavelength of 1.96 μm, the measured photoresponse for the PD also has a good linearity when the input power ranges from 0.17 mW to 1.96 mW, as shown in Fig. 4d.

Figure 4e shows the measured results of the responsivity for the wavelength of 1.55 μm. One sees that the responsivity increases from $7.2\times10^{-5}$ mA/W to 0.147 mA/W as the bias $|V_{bias}|$ increases from 1 V to 6 V. In contrast, when operating at the wavelength of 1.96 μm, the responsivity increases from $3.5\times10^{-5}$ mA/W to 0.0148 mA/W as the bias $|V_{bias}|$ increases from 1 V to 6 V, as shown in Fig. 4f. Even though the optical absorbances for the wavelengths of 1.55 μm and 1.96 μm are similarly as high as ~99.8%, the absorbance-normalized responsivity of the graphene-silicon-graphene PD is very sensitive to the wavelength, which is similar to the silicon-graphene PD discussed above. This also suggests that the PTFE process dominates the photoresponse. Figure 4e-f also shows the NPDR, calculated by NPDR=$R/I_{dark}$ (where $R$ is the responsivity). For the wavelength of 1.55 μm, the NPDR increases from $3.11\times10^3$ W$^{-1}$ to $1.67\times10^6$ W$^{-1}$ as the bias $|V_{bias}|$ increases from 1 V to 2.5 V. As the bias $|V_{bias}|$ further increases to 6 V, the NPDR varies slightly with a peak value of $1.85\times10^6$ W$^{-1}$ at 5.5V. Similarly, the NPDR at 1.96 μm increases from $2.95\times10^3$ W$^{-1}$ to $2.22\times10^5$ W$^{-1}$ as the bias $|V_{bias}|$ increases from 1 V to 2 V. when the bias $|V_{bias}|$ increases further from 2 V to 6 V, the NPDR changes slightly with a peak value of $2.91\times10^5$ W$^{-1}$ at 5.5V (see Fig. 4f).

In order to have a deep look at the PD, here we give an analysis on the band structure of the graphene-silicon-graphene structure qualitatively with a one-dimensional model. As shown in Fig. 5a, a graphene-silicon-graphene structure consists of a pair of con-directional silicon-graphene Schottky junctions (strictly speaking, silicon-SiO$_2$-graphene Schottky junctions). Here the case with a forward bias $V_{bias}$ is considered. In this case, the photoexcited electrons in the FLG at the left and the photoexcited holes in the FLG at the right may contribute to the photocurrent, facing the electron Dirac point barrier $\Phi_{De-IFL}$ and the hole Dirac point barrier $\Phi_{Dh-IFL}$, respectively. Note that $\Phi_{De-IFL}$ is $\Phi_{D-IFL}$ mentioned in Fig. 3. Since the bias voltage for the narrow-sandwiched silicon region is usually negligible,[32] the total bias voltage $V_{bias}$ is given approximately as $|V_{bias}|=|V_{SJ\_L}|+|V_{SJ\_R}|$, where $V_{SJ\_L}$ and $V_{SJ\_R}$ are respectively the reserve bias voltage for the Schottky junction at the left and the forward bias voltage for the Schottky junction at the right. In terms of the bias-polarity definition in this work, in a unidirectional graphene-nSi Schottky diode, the silicon side corresponds to the negative electrode. Most of the electric potential difference is generated at the left Schottky junction, and thus one has $|V_{SJ\_L}|>>|V_{SJ\_R}|$. The voltages can be estimated according to the current continuity condition. The electron current density $J_e$ can be written as[33]

$$J_e(V_{SJ}) = A^{**}T^3 e^{-\chi^{0.5}\delta} \exp[-\frac{\Phi_{Be-IFL}(V_{SJ})}{k_BT}][\exp(\frac{qV_{SJ}}{n_{IF}k_BT})-1], \qquad (Eq.\ 1)$$

where $A^{**}$ is the effective Richardson constant, $T$ is the temperature, $\chi$ is the mean tunneling barrier height presented by the oxide, $\delta$ is the interfacial oxide thickness,[34] $\Phi_{Be-IFL}$ is the Schottky barrier height (written as $\Phi_{B-IFL}$ in Fig. 3 and Fig. S1d), $k_B$ is Boltzmann's constant, $q$ is the elementary charge, and $n_{IF}$ is the diode ideality factor. Here, $n_{IF}$ is about 5.11 according to the analysis of the silicon-graphene I-V curve based on the Cheung method.[35] In a symmetric structure, the current density $J_e$ for the two Schottky junctions are equal, i.e., $|J_e(V_{SJ\_L})|=|J_e(V_{SJ\_R})|$. When $V_{bias}$=6 V, one has $V_{SJ\_L}$=−4.59 V and $V_{SJ\_R}$=1.41 V, while the depletion widths are respectively 256 nm and 104 nm based on this 1D model. When simplifying the actual graphene-silicon-graphene structure with a coplanar stripline configuration[36] to the 1D model, it is well known that the equivalent separation between the two Schottky junctions (Fig. 5a) is larger than the graphene gap $W_{gap}$ (i.e., 300 nm here). Consequently, silicon should be nearly completely depleted for the present case according to the 1D model estimation given above.

Figure 5b exhibits the results of the simulated barriers, where $\Phi_{De}$ and $\Phi_{De-IFL}$ are obtained directly from the modeling results in **Supplementary Note 3**. The Dirac point barrier height $\Phi_{Dh}$ (ignoring the IFL effect) is given by $\Phi_{Dh}=E_g-\Phi_{De}$, where $E_g$ is the silicon bandgap of 1.12eV[37]. As a result, $\Phi_{Dh-IFL}$ can be calculated by Eq. S7 (see the details in **Supplementary Note 3**), while the IFL effect on the hole barrier should be considered only when the silicon built-in potential $\varphi_i$<0. Correspondingly, one has $V_{SJ}$ >0.56 V in this case (see region III in Fig. 5b). Correspondingly, $\Phi_{De-IFL}$ is calculated only when $\varphi_i$>0, where the electron barriers are calculated for the left reverse barrier with $V_{SJ}$<0 (see region I in Fig. 5b). As shown in Fig. 5b, the two Schottky junctions have $\Phi_{De-IFL}$=0.363 eV (@ $V_{SJ\_L}$=−4.59 V) and $\Phi_{Dh-IFL}$ =0.479 eV (@ $V_{SJ\_R}$= 1.41 V), respectively. As shown in Fig. S1d, the chemical potentials $\mu_g$ for the graphene sheets at the left and right sides are respectively −0.129 eV and −0.17 eV, indicating that both parts of FLGs have full optical absorption ability. Similarly to the silicon-graphene PD, the photoresponse mechanisms for the graphene-silicon junction at the left is respectively the IPE and PTFE processes. For the graphene-silicon junction at the right, half of the photon energy ($h\nu/2$) are still below $\Phi_{De-IFL}$ and $\Phi_{Dh-IFL}$, the PTFE mechanism is the dominating mechanism at both wavelengths of 1.55 μm and 1.96 μm. Since $\Phi_{De-IFL}$<$\Phi_{Dh-IFL}$, the photoexcited electron current is the main component here. It should be noted that the one-dimensional model does not give a precise modelling but nevertheless provides a simple method to reveal how the photocurrent generates.

Figure 6a shows the experimental setup for the high frequency measurement, including a commercial modulator and a vector network analyzer (VNA) (see **Materials and Methods)**. Figure 6b gives the normalized high frequency response of the graphene-silicon-graphene PDs. The measured 3 dB bandwidth $f_{3dB}$ is about 33 GHz, which is owing to the small RC time constant as well as the short carrier transit time in the present PD. Here the RC time constant is small due to low graphene-metal contact resistance and low capacitance of the active region. Furthermore, the total carrier transit time is short because of the short transit time in silicon and the fast carrier dynamics at the silicon-graphene interface.

In this work, we fabricated several graphene-silicon-graphene waveguide PDs and the measured results are summarized in **Supplementary Note 3**. At the wavelength of 1.55 μm, the responsivity is 0.13-0.33 mA/W, while the directly measured bandwidth from the VNA is 19-33 GHz. And the NPDRs are $1.5 \times 10^6$-$6.5 \times 10^6$ W$^{-1}$. In addition, these devices exhibit large linear optical power ranges of 3 μW-1.86 mW (and more) at 1.55 μm and 0.17-1.96 mW (and more) at 1.96 μm, respectively (see **Table S2**).

**Discussions and perspective.** In graphene-silicon-graphene PDs, high energy barrier faced by the photo-excited holes may lead to a low IQE compared to silicon-graphene PDs in theory. Interestingly, the present graphene-silicon-graphene PDs exhibits higher responsivity and higher bandwidths than the silicon-graphene PDs[30] in our experiments. The performance of our fabricated silicon-graphene PD in ref. [30] may be limited by the imperfection Al-Si contact with a high contact resistance and a high Schottky barrier. In contrast, there is no metal-silicon contact introduced for the graphene-silicon-graphene PDs presented in this paper.

In **Supplementary Note 4**, we give a summary for the performances of the representative waveguide-integrated Si-2DM PDs. As shown in **Table S3**, the graphene bolometers with the metal-graphene-metal structure can operate with a high bandwidth but a very large dark current, which results in low NPDRs of 10-$10^2$ W$^{-1}$ (see **Table S3**). Besides, graphene bolometers usually have poor linearity performances.[25,38] The metal-2DM-metal PDs have improved NPDRs when using non-zero bandgap 2DMs instead of graphene, as shown in **Table S3**. However, currently it has never achieved a large NPDR over $10^5$ W$^{-1}$ and a large bandwidth over 1 GHz simultaneously.[39-42] Alternatively, a graphene p-n homojunction PD is interesting because it possibly with zero-bias operation, in which case the dark current and the dark-current shot noise is zero (but the thermal noise remains). And the bandwidth is possible as high as >67 GHz as the stat-of-the-art.[6] Another ring-resonator PD exhibited a bandwidth of 12 GHz and a high responsivity of 90 V/W,[7] resulting in a high sensitivity on par with Ge/Si PDs. However, its LDR is limited to be 0.01-0.2 mW. More recently, an Au-MoS$_2$ Schottky PD using an Au active region exhibited a responsivity of 15.7 mA/W, an NPDR of $1.05 \times 10^4$ W$^{-1}$, and a bandwidth of 1.37 GHz at 1.55 μm.[43]

The 2DM heterostructure PDs have shown the potential to feature overall high performances. For instance, a MoTe$_2$-graphene heterostructure PD has an NPDR of $10^4$-$10^6$ W$^{-1}$ and bandwidths of 12-46 GHz at the wavelength band of 1.31 μm. Another MoS$_2$/G-hBN-G heterostructure PD was reported with a high NPDR of $10^7$ W$^{-1}$ and a bandwidth of 28 GHz at 1.55 μm. However, the linearity performance is unknown yet. In contrast, the silicon-graphene heterostructure PD based on the wet-transferred CVD-grown graphene have the advantages of easy fabrication and high feasibility of large-scale photonic integration, as discussed in **Supplementary Note 4**. Currently very few waveguide-integrated silicon-graphene heterostructure PDs have been reported. An Au-silicon-graphene heterostructure PD has shown a relatively high responsivity of 85 mA/W.[13] Another waveguide-integrated silicon-graphene p-i-n photodiode for the 1.55 μm light exhibited a responsivity of 3.4 mA/W at zero bias and a high RF bandwidth estimated from the converted photoelectric signal with a center frequency of 40 GHz.[14] In the present work, several graphene-silicon-graphene heterostructure PDs are demonstrated with a responsivity of 0.13-0.33 mA/W, a bandwidth of 19-33 GHz, and a large NPDR of $1.5\times10^6$-$6.5\times10^6$ W$^{-1}$. Our PDs exhibit large LDRs of 3 μW-1.86 mW (and beyond) at 1.55 μm and 0.17-1.96 mW (and beyond) at 1.96 μm, respectively (see **Table S2**).

According to the comparison shown in **Table S3**, the present PD shows excellent overall performances of the NPDR, the bandwidth, and the linearity among the counterparts. Moreover, the fabrication-friendly graphene-silicon-graphene heterostructure PDs have excellent reproductivity and stability. Apparently, for the present silicon-graphene heterostructure PDs, one of the most important works in the future is to further improve the quantum efficiency. More efforts should be made for the tunneling current modeling for IQE evaluations. Here the reveal of the photoresponse mechanisms at the silicon-graphene interface provides the potential solution for the performance improvements. For example, the doping engineering of silicon may be helpful. Heavier doping in silicon can reduce the depletion region width[37] and further reduce the tunneling length for photoexcited carriers, which is possibly to significantly improve the IQE. On the other hand, the dark current would probably increase simultaneously. Fortunately, the NPDR performance would be still very acceptable.

**CONCLUSIONS**

In this work, we have demonstrated graphene-silicon-graphene heterostructure waveguide PDs for the wavelength-bands of 1.55/2 μm. The fabricated PDs have exhibited excellent linearity in the power range of 0.17-1.96 mW and a high NPDR of $2.66\times10^5$ W$^{-1}$ when operating at 1.96 μm. At the wavelength of 1.55 μm, the present PDs also show excellent linearity in the power range of 3 μW-1.86 mW (limited by the maximal power coupled to the chip) and a high NPDR of $1.63\times10^6$ W$^{-1}$. The measured 3-dB bandwidth is as high as ~33 GHz, which is attributed to the small RC constant, the short transit time in silicon, as well as the fast carrier dynamics in the silicon-graphene junction. On the other hand, the PDs still have limited responsivities of ~0.15 and ~0.015 mA/W for the wavelength-bands of 1.55 μm and 2 μm, respectively. According to the theoretical model established for the silicon-graphene heterostructure, the dominant mechanisms for the photoresponse of the present PDs include the IPE and the PTFE effect, and thus it is possible to improve the responsivity performance by utilizing some strategies such as doping engineering.[37] More efforts should be made to develop graphene-silicon-graphene waveguide PDs with high overall-performances to satisfy the demands for large-scale photonic integration in the future.

## MATERIALS AND METHODS

**Device Fabrication.** A silicon-on-insulator (SOI) wafer with a 100-nm-thick top-silicon layer was obtained from a 220-nm-thick SOI wafer (p-type with a doping concentration of $10^{14}$ cm$^{-3}$) by the processes of thermal oxidation and wet etching. The *n*-doped silicon was formed with an ion implantation process. The processes of EBL and ICP were then carried for the fabrication of silicon ridge waveguides. The buffered-oxide-etch (BOE) cleaning was carried, and the 50/35-nm Al/Au electrodes was fabricated by using the processes of electron beam evaporation, EBL, and lift-off. The Al/Au electrodes were used as the bottom layer of the sandwiched metal-graphene-metal electrodes. Then, graphene was transferred and patterned by the processes of EBL and ICP. Finally, a 40-nm-thick Au layer was deposited and patterned to form the top layer of the metal-graphene-metal electrodes.

**Transfer process of graphene.** The CVD-grown graphene was obtained from SUNANO Group (few layers, on copper foil). A 300-nm-thick film of PMMA was spin-coated on the graphene/copper film at 4000 rpm. The PMMA/graphene/copper film was floated on aqueous ammonium persulfate (60 mg/mL) to remove the copper and rinsed in deionized water. Then, the film was transferred onto the chip. The graphene-covered chip was dried, baked, soaked in acetone and rinsed with isopropanol.

**Device measurement.** The static I-V curves of the PDs were measured by a source meter for different injected optical powers. Light from a laser (a tunable laser @ 1.55 μm or a fiber laser @ ~2 μm) was polarization-controlled and then coupled to the on-chip waveguide with the corresponding grating coupler. The photocurrent was then calculated by subtracting the dark current from the total current. The input optical power $P_{in}$ was estimated according to the measured coupling efficiency of the grating coupler and the excess loss of the wavelength multiplexer. More details on the optical power analysis are provided in **Supplementary Note 1**. In the high frequency measurement, the CW light was modulated with a commercial optical modulator (Sumitomo T. MXH1.5DP-40PD-ADC, 22 GHz bandwidth) by the RF signal from a vector network analyzer

(ROHDE & SCHWARZ ZVA40, 40 GHz). The modulated light was amplified by an EDFA (Thorlabs EDFA100P) before it was coupled to the chip. The output electrical signal of the test device was then amplified by using a RF amplifier (Centellax OA4MVM2) and finally received by the VNA.

## ASSOCIATED CONTENT

**Supporting Information Available:** The optical power calibration; Modeling of the graphene-silicon Schottky junction; Performance summary of the waveguide-integrated graphene-silicon-graphene heterostructure PDs; Performance summary of the waveguide-integrated Si-2DM PDs working at the NIR/MIR ranges.

AUTHOR INFORMATION

**Corresponding Author**

Daoxin Dai − State Key Laboratory for Modern Optical Instrumentation, Zhejiang Provincial Key Laboratory for Sensing Technologies, College of Optical Science and Engineering, International Research Center for Advanced Photonics, Zhejiang University, Zijingang Campus, Hangzhou, 310058, China.; Intelligent Optics & Photonics Research Center, Jiaxing Research Institute, Zhejiang University, Jiaxing 314000, China; Ningbo Research Institute, Zhejiang University, Ningbo 315100, China;

Email: dxdai@zju.edu.cn

**Author Contributions**

J. Guo and C. Liu designed the devices. C. Liu performed the device fabrication with the assistance of L. Yu. C. Liu performed the device measurements with the assistance of J. Guo and L. Yu. J. Guo performed the simulations and modeling. J. Guo, C. Liu and D. Dai analyzed the simulation and experimental results. J. Guo and D. Dai wrote the manuscript with the assistance of C. Liu and H. Xiang. All of the authors contributed to the discussions and the manuscript revisions. D. Dai supervised the project. [#]J. Guo and [#]C. Liu contributed equally.

**Funding Sources**

This work was supported by National Major Research and Development Program (No. 2018YFB2200200/2018YFB2200203); National Science Fund for Distinguished Young Scholars (61725503); National Natural Science Foundation of China (NSFC) (61905210, 61961146003, 91950205); Zhejiang Provincial Natural Science Foundation (LR22F050001, LD22F040004); The Fundamental Research Funds for the Central Universities.

**Notes**

The authors declare no conflict of interest.

**ACKNOWLEDGMENT**

We thank ZJU Micro-Nano Fabrication Center for help on the device fabrication. We thank Maopeng Xu and Prof. Yizheng Jin for performing the KPFM measurement.

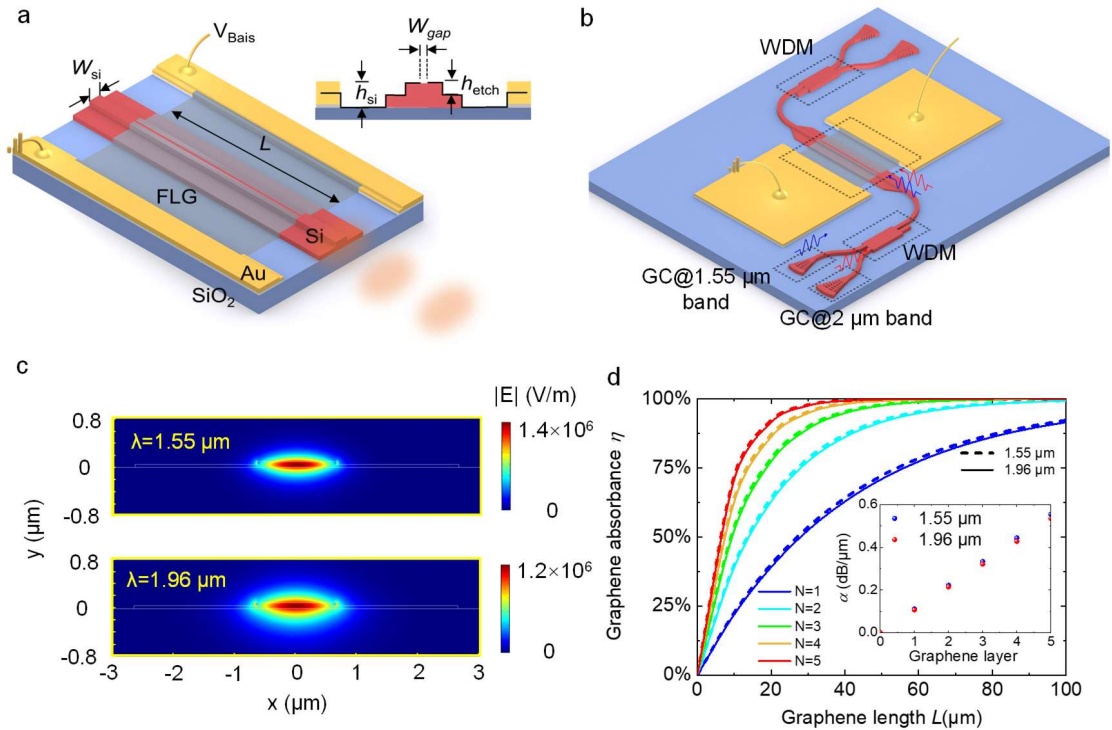

**Figure 1.** The graphene-silicon-graphene waveguide PDs. **a-b** 3D schematic diagram and the cross-section of the graphene-silicon-graphene PDs **a** and the photonic integrated circuit for testing **b**. **c** The mode fields of the thin-silicon/graphene waveguide at the wavelengths of 1.55/2 μm. Here four-layer graphene is considered. **d** The optical absorbance at 1.55/1.96 μm for the cases with different graphene layer numbers $N$=1, 2, 3, 4, and 5. Inset: the mode absorption coefficient $\alpha$ as a function of the graphene layer number $N$. The parameters are $W_{Si}$=1.3 μm, $W_{gap}$=300 nm, $h_{Si}$=100 nm, and $h_{etch}$=50 nm. FLG: few-layer graphene, λ: wavelength. WDM: wavelength-division (de)multiplexer. GC: grating coupler.

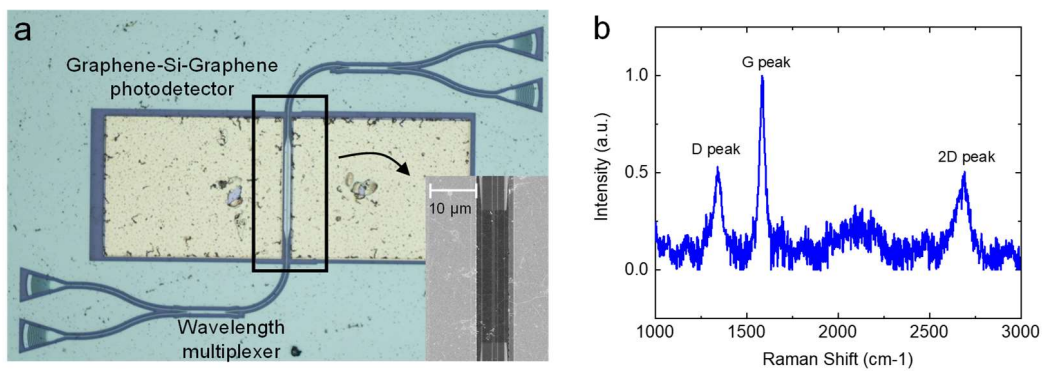

**Figure 2.** Characterization of the fabricated sample. **a** Optical microscopy picture of the fabricated PIC. Inset: SEM picture of the 30-μm-long FLG-Si-FLG waveguide PD. **b** Normalized Raman spectrum of the FLG used for the PD.

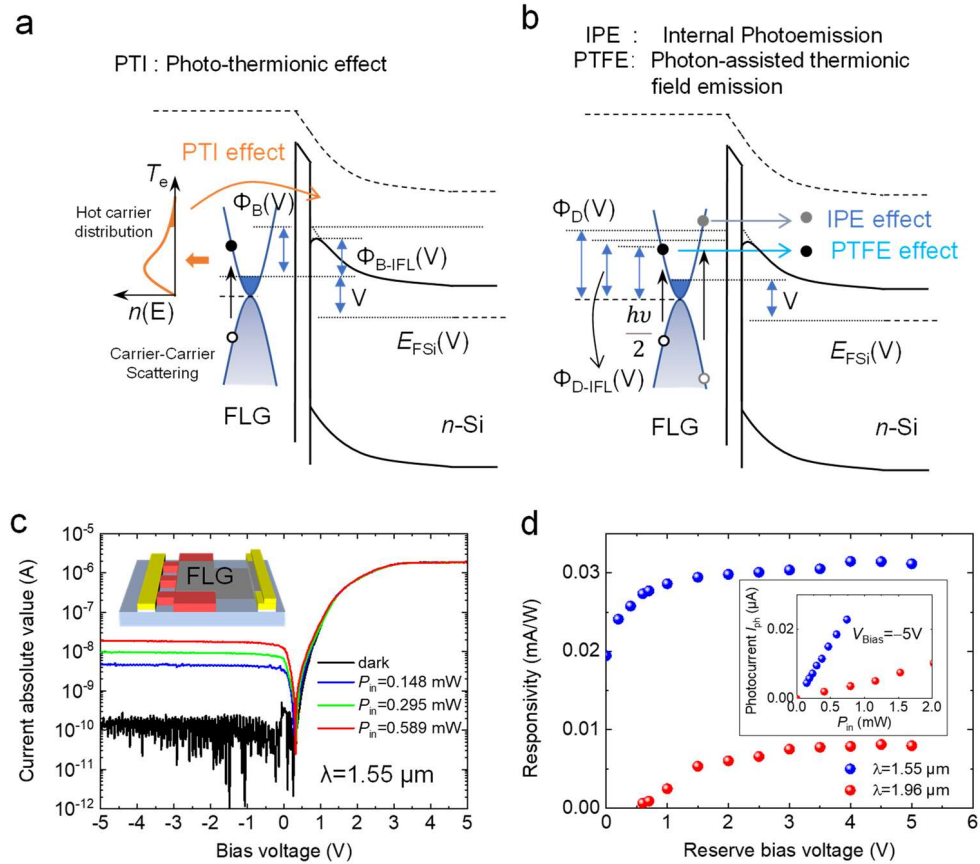

**Figure 3.** The silicon-graphene Schottky PD. **a-b** Band-diagrams for two types of photoresponse mechanisms: **a** the PIT effect, where $n(E)$ is the carrier distribution; **b** the direct photodetection effects, including the IPE and the PTFE. **c**. The measured I-V curves at 1.55 μm. **d** The measured responsivities as functions of the reserve bias for the wavelengths of 1.55 μm and 1.96 μm. Inset: the relationship of $I_{ph} \sim P_{in}$ at $V_{bias}=-5V$.

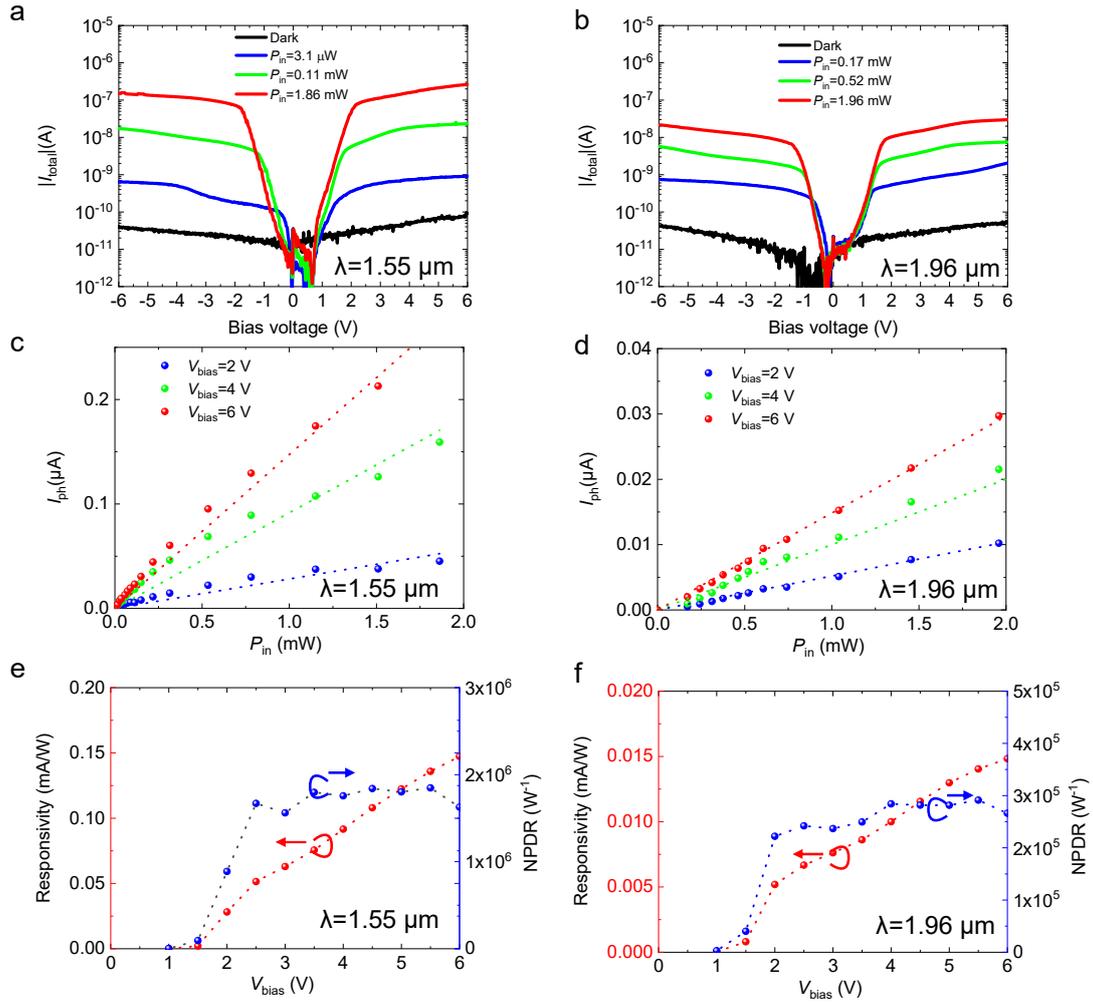

**Figure 4.** The measured results of a typical graphene-silicon-graphene device (Device A), **a-b** the I-V curves at 1.55 μm **a** and 1.96 μm **b**, **c-d** the results of $I_{ph} \sim P_{in}$ (dotted line: the fitting curves) at 1.55 μm **c** and 2 μm **d**, **e-f** The responsivity and the NPDR as a function of $V_{bais}$ at 1.55 μm **e** and 1.96 μm **f**.

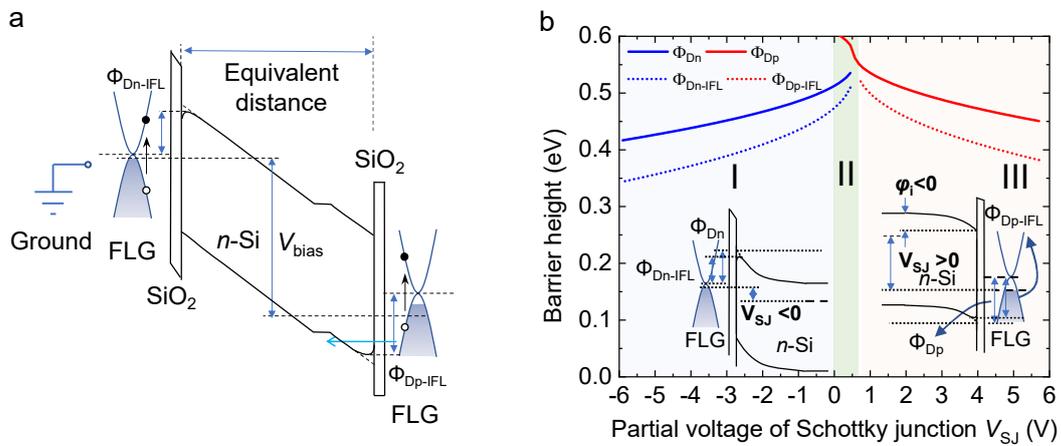

**Figure 5. a** Band diagram of the graphene-silicon-graphene structure consisting of a pair of condirectional Schottky junctions. **b** The barrier heights of a silicon-graphene Schottky junction as a function of the bias voltage $V_{SJ}$, inset: the band diagrams, region I: $V_{SJ}<0$, region II: $V_{SJ}<0$, silicon built-in potential $\varphi_i>0$, region III: $V_{SJ}<0$, silicon built-in potential $\varphi_i<0$. The graphene layer number $N=4$, Si is *n*-doped with the doping concentration of $10^{17}$ cm$^{-3}$. $I_{ph}$: the photocurrent; $P_{in}$: the input optical power; $V_{bias}$: bias voltage of the graphene-silicon-graphene PD. $V_{SJ}$: partial voltage of the silicon-graphene Schottky junction.

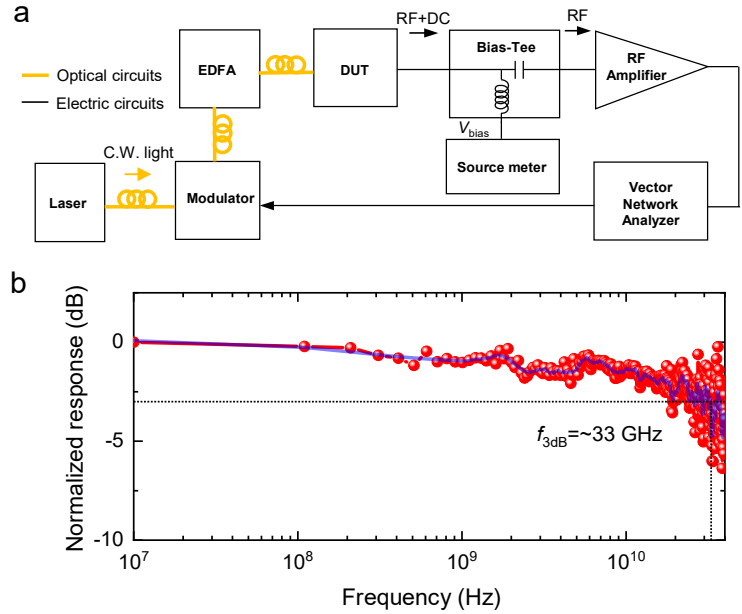

**Figure 6.** The high frequency measurement. **a** setup, **b** normalized response of Device A at 1.55 μm. Red dot line: origin data, blue line: data after noise reduction for the bandwidth evaluation.

# Supplementary Information

# High-speed graphene-silicon-graphene waveguide photodetectors with high photo-to-dark-current ratio and large linear dynamic range

Jingshu Guo[†,‡,#], Chaoyue Liu[†,#], Laiwen Yu[†], Hengtai Xiang[†], Yuluan Xiang[†], and Daoxin Dai[*,†,‡,§]

† State Key Laboratory for Modern Optical Instrumentation, Zhejiang Provincial Key Laboratory for Sensing Technologies, College of Optical Science and Engineering, International Research Center for Advanced Photonics, Zhejiang University, Zijingang Campus, Hangzhou, 310058, China.

‡ Intelligent Optics & Photonics Research Center, Jiaxing Research Institute, Zhejiang University, Jiaxing 314000, China.

§ Ningbo Research Institute, Zhejiang University, Ningbo 315100, China.

# These authors contributed equally to this work: Jingshu Guo, Chaoyue Liu.

*Corresponding author e-mail: dxdai@zju.edu.cn.

**Supplementary Note 1. The optical power calibration.**

As shown in Fig. 1b, the input power $P_{in}$ is extracted by $P_{in} = P_{inchip} - (EL_{GC} + EL_{WDM})$, where $P_{inchip}$ is the measured optical power to be coupled to the chip in dBm, $EL_{GC}$ and $EL_{WDM}$ are respectively the excess losses of the grating coupler (GC) and wavelength-division (de)multiplexer (WDM) in dB. $EL_{GC}$ was measured by the testing structure consisting of a pair of grating couplers. According to the results from multiple samples, we have $EL_{GC}$ (λ @ 1.55 μm)=6.8±0.28 dB and $EL_{GC}$ (λ @ 1.96 μm)=8.0±0.2 dB. The total loss of ($EL_{GC}+EL_{WDM}$) was measured by characterizing the testing structure consisting of a pair of GCs and a pair of WDMs. The total loss of ($EL_{GC}+EL_{WDM}$) is about 8.6±1 dB at 1.55 μm and 9.15±0.13 dB at 1.96 μm, respectively.

**Supplementary Note 2. Modeling of the graphene-silicon Schottky junction**

The modeling of the graphene-silicon Schottky junction is given to have a quantitative analysis of the heterostructure band diagram, which helps clarify the operation mechanism of the photo-assisted carrier dynamics. A significant difference between the metal-silicon Schottky junction and graphene-silicon Schottky junction is that the limited density of states near the Dirac point makes the graphene chemical potential varied as the graphene-doping-induced barrier lowering.[1,2] In Table S1, the definitions and the values of the parameters are listed. In the modeling, the temperature $T$ is 300 K, and the $n$-Si has a doping concentration of $N_{Si}=10^{17}$ cm$^{-3}$, and $\varphi_n$= 0.145 eV.[3]

Table S1 The parameter table for graphene-Si Schottky junction modeling.

| Parameter | Definition | Value | Unit |
|---|---|---|---|
| T | Temperature | 300 | K |
| $E_{FSi}$ | Femi level of Si | - | eV |
| $E_g$ | Band gap of Si | 1.12 | eV |
| χ | Electron affinity of Si | 4.05 | eV |
| $\varphi_i$ | Built-in potential in Si | - | V |
| $N_{si}$ | Doping concentration in Si | $10^{17}$ | cm$^{-3}$ |
| $\varphi_n$ | The energy difference between the conduction band bottom and the Fermi level of n-doping Si | 0.145[3] | eV |
| $Q_{Si}$ | Space charge density in Si | - | C·m$^{-2}$ |
| $\varepsilon_0$ | Permittivity of vacuum | 8.85×10⁻¹² | F/m |
| $\varepsilon_{Si}$ | Relative permittivity of Si | 11.9 | - |
| $\varepsilon_{sio2}$ | Relative permittivity of SiO$_2$ | 3.9 | - |
| $\varphi_o$ | Potential difference across the SiO$_2$ oxide layer | - | eV |
| $t_o$ | Thickness of the SiO$_2$ oxide layer | 1 | nm |
| $\Phi_G$ | The work function of intrinsic graphene denoting the potential difference between the Dirac point of graphene and the vacuum level | 4.6[4] | eV |
| $\Phi_B$ | Barrier height with the image force lowering ignored | - | eV |
| $\Phi_{B-IFL}$ | Barrier height with the image force lowering considered | - | eV |
| $\Phi_D$ | The potential difference between the Dirac point of graphene and the highest conduction band bottom energy of Si | | |
| $\Phi_{D-IFL}$ | $\Phi_D$ with the image force lowering considered | - | eV |
| $\mu_g$ | The chemical potential of graphene | - | eV |
| $n_g$ | The positive carrier density in few-layer graphene | | |
| $E_{FG}$ | Fermi level of graphene | - | eV |
| $v_F$ | Graphene Fermi velocity | 10⁶ | m/s |

Notes: The subscript '0' denotes the variable at zero bias, and the suffix '(V)' denotes the variable under bias.

Fig. S1a shows the band diagram under zero bias. In the graphene-Si contact with the space charges, graphene chemical potential is $\mu_{g0}$ (Fig. S1a). Here, the positive and negative chemical potentials correspond to the cases of electron doping and hole doping, respectively. The barrier height $\Phi_{B0}$ at zero bias is set as a variable. The built-in potential in silicon is

$$\varphi_{i0} = \Phi_{B0} - \varphi_n. \quad \text{(Eq. S1)}$$

The space charge density $Q_{Si0}$ is

$$Q_{Si0} = \text{sgn}(\varphi_{i0} - k_B T/q)\sqrt{2q\varepsilon_0\varepsilon_{Si}N_{Si}|\varphi_{i0} - k_B T/q|}. \quad \text{(Eq. S2)}$$

The depletion width in silicon is estimated from $W_{\text{depletion}} = |Q_{Si}/qN_{Si}|$.[3] The potential across the SiO$_2$ oxide layer is then extracted by

$$\varphi_{o0} = q \cdot Q_{Si}t_o/\varepsilon_0\varepsilon_{sio2}. \quad \text{(Eq. S3)}$$

Then, for $n$-doping silicon, the graphene chemical potential is given by (see Fig. S1a)

$$\mu_{g0} = \Phi_G - \varphi_{o0} - \chi - \Phi_{B0}. \quad \text{(Eq. S4)}$$

Apparently, the Dirac point barrier height can be given by

$$\Phi_{D0} = \mu_{g0} + \Phi_{B0}. \quad \text{(Eq. S5)}$$

Using Eqs. S1-S4, the chemical potential $\mu_{g0}$ is given as a function of $\Phi_{B0}$. To obtain $\mu_{g0}$, the Kelvin probe force microscopy (KPFM) measurement was carried by using highly oriented pyrolytic graphite (HOPG) as calibration. The FLG on silicon ($n$-doping, 10¹⁷ cm⁻³) has a work function of 4.75 eV and the work function $\Phi_G$ of the intrinsic graphene is 4.6 eV,[4] thus the chemical potentials is −0.15 eV. Correspondingly, $\Phi_{B0}$ is estimated to be 0.663 eV.

The band structures of the FLG with various stacking types have been researched early.[5,6] Ignoring the interlayer neighbor hopping, the sub-band dispersion relations $E\sim k_\parallel$ of the FLG are assumed to be the same as that of the single-layer graphene (SLG). With this assumption, the density of states (DOS) is proportional to the layer number, and therefore the positive charge density of the $N$-layer FLG $n_g$ can be expressed by $N$-folds of SLG positive charge density $n_{SLG}$.

$$n_g \approx n_{SLG} \cdot N \approx -\text{sgn}(\mu_g)\frac{|\mu_g|^2}{\pi(\hbar v_F)^2} \cdot N = f(\mu_g) \quad \text{(Eq. S6)}$$

In this work, one has $N=4$. It should be noted that the expression for the carrier density $n_{SLG}$ is valid when $T=0$ K and established approximately at room temperature (i.e., 300 K here). For the FLG, it is assumed that there is a linear relationship between DOS and $N$.[7] It should be noticed that the image force lowering (IFL) effect is an important issue that needs to be considered when $\varphi_i>0$, as shown in Fig. S1a. The image force lowering amount $\triangle\Phi_{IFL}$ can be calculated by the built-in potential of silicon.[8]

$$\Delta\Phi_{IFL} = \left[\frac{q^3 N_{si} \frac{\varphi_i(V)}{q}}{8\pi^2(\varepsilon_0\varepsilon_{Si})^3}\right]^{1/4} \cdot q. \tag{Eq. S7}$$

Apparently, one has $\Phi_{B-IFL}=\Phi_B-\triangle\Phi_{IFL}$ and $\Phi_{D-IFL}=\Phi_D-\triangle\Phi_{IFL}$. All the band diagram parameters under zero bias have been extracted now.

For the bias condition, the subscripts '0' is replaced by the suffixes '(V)'. As shown by the band diagrams in Fig. S1b and c, a forward/reverse bias induces less/more negative doping charges in graphene. The built in potential $\varphi_i(V)$ in the silicon is given as $\varphi_i(V) = \varphi_{i0} + \Delta\varphi_i(V)$ at any given biases. As a result, one can obtain the space charge density $Q_{si}(V)$ by Eq. S2, and get the potential $\varphi_o(V)$ across the SiO$_2$ oxide layer by Eq. S3. Then the graphene charge density is given as $n_g(V)=n_{g0}-[Q_{Si}(V)-Q_{si0}]/q$, where the graphene chemical potential is given $\mu_g(V) = f^{-1}[n_g(V)]$. Now, the barrier height at the bias voltage $V$ can be calculated by

$$\Phi_B(V) = \Phi_{B0} - [\varphi_o(V) - \varphi_{o0}] - [\mu_g(V) - \mu_{g0}]. \tag{Eq. S8}$$

The Dirac point barrier height $\Phi_D(V)$ can be extracted by Eq. S6. Considering the IFL effect, the barrier height $\Phi_{B-IFL}(V)$ and $\Phi_{D-IFL}(V)$ can be calculated with Eq. S7. As shown in Fig. S1b, c, the bias voltage is extracted by

$$V \cdot q = -[\mu_g(V) - \mu_{g0}] - [\varphi_i(V) - \varphi_{i0}] - [\varphi_o(V) - \varphi_{o0}]. \tag{Eq. S9}$$

In this way, we obtain the barrier heights and the bias voltage as the built in potential $\varphi_i(V)$ varies. In other words, we have extracted the variables (i.e., $\Phi_B$, $\Phi_{B-IFL}$, $\Phi_D$, $\Phi_{D-IFL}$, $\mu_g$, $\varphi_i$) as functions of the bias voltage. The calculated barrier heights are given in Fig. S1d. When $\varphi_i<0$, the bias voltage V>0.56 V here, and the IFL effect can be ignored.[3] $\mu_g$ is $-0.124\sim-0.185$ eV at the bias voltage of $-6\sim5.71$V, indicating that the variation of the graphene chemical potential has little influence on the light absorption in graphene.[9] Indeed, we found that the total chip losses remain the same almost as the reserve bias voltage increases in experiments. As the bias voltage change from 0 to $-6$ V, the barrier height $\Phi_{B-IFL}$ decreases from 0.625 eV to 0.467 eV, and the barrier height $\Phi_{D-IFL}$ decreases from 0.475 eV to 0.343 eV.

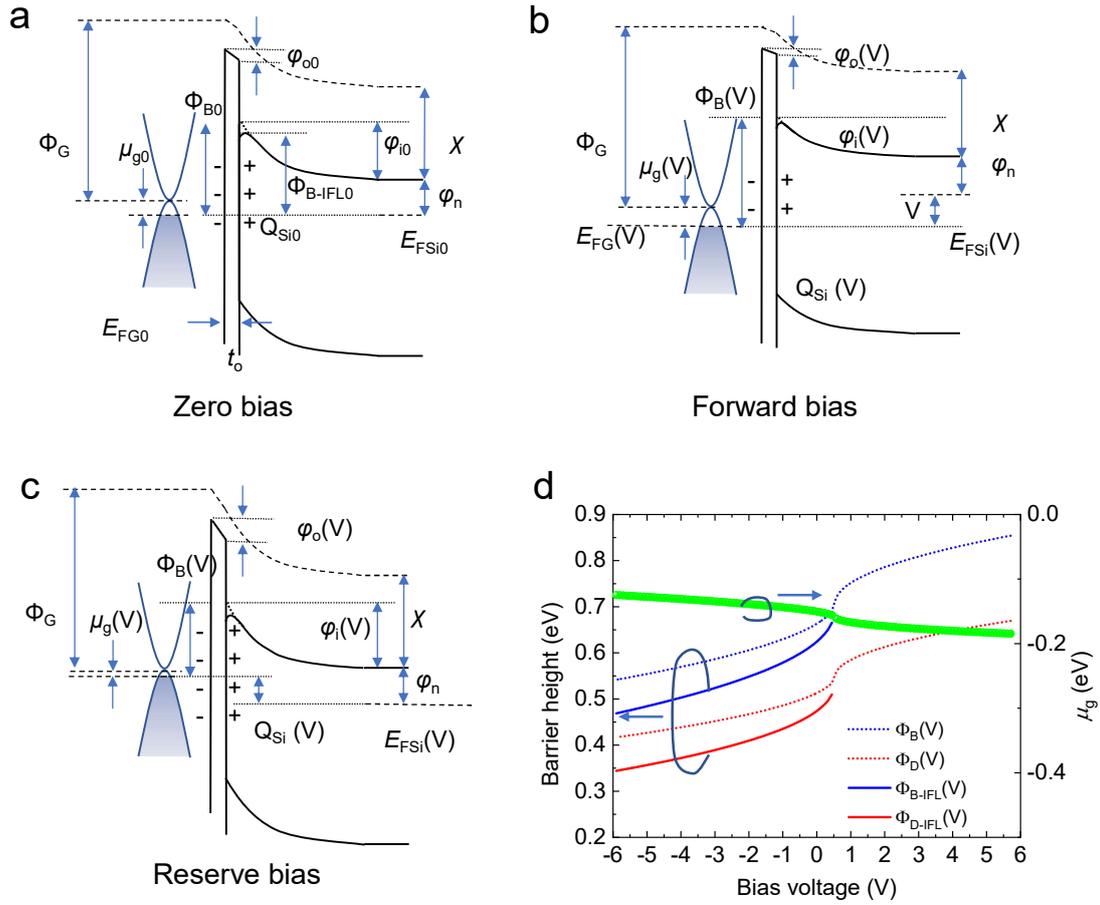

**Fig. S1 Modeling of the graphene-silicon Schottky junction. a-c** Band diagram of the graphene-silicon Schottky junction at zero bias **a**, forward bias **b**, and reserve bias **c**. **d** The calculated barrier heights and the graphene chemical potential as functions of the bias voltage, $\Phi_{B0}$ is assumed to be 0.663eV. The $n$-Si has a doping concentration of $10^{17}$ cm$^{-3}$. The graphene layer number is 4.

**Supplementary Note 3. Performance summary of the waveguide-integrated graphene-silicon-graphene heterostructure PDs.**

In this work, we have fabricated several PDs with various active region lengths. In Table S2, the performances of these PDs are summarized and demonstrated, including Device A introduced in main text. Although the lengths of the graphene sheets range from 30 μm to 80 μm, the high mode absorption coefficient of ~0.44 dB/μm (4-layer graphene) lead to a nearly saturated absorbance of >95%. Devices A-E were fabricated on the $n$-Si chip with a doping concentration of $10^{17}$ cm$^{-3}$, while Device F was based on $p$-Si chip with a doping concentration of $10^{14}$ cm$^{-3}$. The responsivities at 1.55 μm are 0.147~0.326 mA/W, which are mainly about 10-20 folds of the responsivities at 1.96 μm. For Device F, the responsivity at 1.55 μm is ~50 folds higher than that at 1.96 μm. The maximal responsivity was measured from Device C and one has 0.326 mA/W at 1.55 μm under 5V bias. The devices have high NPDRs in the scale of $10^6$ W$^{-1}$ at 1.55 μm. At 1.96 μm, the NPDR is in the scale of $10^5$ W$^{-1}$. The bandwidths are estimated to be independent of the measured wavelength. The measurement results show the bandwidth at 1.55 μm is 18.8-33 GHz. In general, our PDs have shown fairish reproductivity.

**Table S2** Performance summary of the as-fabricated waveguide-integrated G-Si-G PDs.

| Device | $L_{gra}$(μm) | λ(μm) | Static performance | | | Bandwidth (GHz) | Si doping |
|---|---|---|---|---|---|---|---|
| | | | Responsivity (mA/W) | NPDR($W^{-1}$) | Bias(V) | | |
| A | 50 | 1.55 | 0.147 | $1.63\times10^6$ | 6 | 33 GHz | N type, $10^{17}$ $cm^{-3}$ |
| | | 1.96 | 0.015 | $2.66\times10^5$ | 6 | - | |
| B | 30 | 1.55 | 0.171 | $5.76\times10^6$ | 6 | 18.8 GHz | |
| | | 1.96 | 0.017 | $3.79\times10^5$ | 6 | - | |
| C | 50 | 1.55 | 0.326 | $6.52\times10^6$ | 5 | ~20 GHz | |
| D | 60 | 1.55 | 0.132 | $1.84\times10^6$ | 5 | 27 GHz | |
| | | 1.96 | 0.013 | $1.93\times10^5$ | 5 | - | |
| E | 80 | 1.55 | 0.156 | $1.54\times10^6$ | 5 | 25 GHz | |
| | | 1.96 | 0.007 | $3.92\times10^4$ | 5 | - | |
| F | 80 | 1.55 | 0.193 | $6.43\times10^6$ | 5 | 27 GHz | P type, $10^{14}$ $cm^{-3}$ |
| | | 1.96 | 0.004 | $1.15\times10^5$ | 5 | - | |

**Supplementary Note 4. Performance summary of the waveguide-integrated Si-2DM PDs working at the NIR/MIR ranges.**

Here we give a summary for the performance of the waveguide-integrated Si-2DM PDs working at near-infrared/mid-infrared ranges, as shown in Table S3. Previously, we have classified the Si/2DM PDs by three types of configurations, namely, metal-2DM-metal, metal-2DM+X-metal, 2DM-Heterostructure.[10] Currently, the waveguide-integrated Si-2DM PDs are mainly based on the metal-2DM-metal and 2DM-heterostructure configurations. Many metal-2DM-metal PDs work based on the bolometric effect,[11,12] the photovoltaic effect in the photoconductive mode,[13-15] or the photoconductive effect[16] under bias. The graphene p-n homojunction PDs can work under zero bias by utilizing the photo-thermoelectric (PTE) effect.[17,18] As a special case. the Au-$MoS_2$ PD can work based on the IPE effect at 1.55 μm (beyond the cut-off wavelength edge of $MoS_2$).[19] The 2DM-heterostructure PDs are usually based on the 2DM-2DM heterostructure[20,21] and the G-Si heterostructure.[22,23]

To make the comparison as comprehensive as possible, the reproductivity is shown by listing the 2DM type.[24] Generally speaking, the CVD-grown 2DMs based on the wet transfer method [11,12,14,21-23] or the semi-dry transfer method[17] are suitable for large-scale photonic integration, compared to the mechanical exfoliation method. In addition, less types of 2DM can greatly reduce the process difficulty. As a result, the G-Si PDs[22,23] still have unique advantages in scalability and integration. The PDs should be evaluated from the performances, including the operation wavelength, the responsivity, the linear dynamic range (LDR), the NPDR, and the bandwidth.

For PDs, the NEP (noise equivalent power) is a strict indicator to evaluate the sensitivity, considering the total noise including the thermal noise, the shot noise, the 1/$f$ noise, and the generation-recombination ($g$-$r$) noise. In practice, the total noise is not easy to evaluate strictly.[25] Alternatively, the NPDR (=$R/I_{dark}$) is an intuitive indicator for sensitivity evaluation. In terms of the bandwidth, the results for the time-domain pulse measurement and the Fourier transform are marked.

**Table S3** Performance comparison among the reported waveguide-integrated Si-2DM

photodetectors working at the NIR/MIR wavelengths.

| Type | Structure | 2DM Type | λ(μm) | Responsivity | Bias | linear dynamic range[a] | NPDR (W$^{-1}$)[b] | Bandwidth[c] | Refs. |
|---|---|---|---|---|---|---|---|---|---|
| metal-2DM-metal | M-G-M | CVD | ~1.55 | 200-400 mA/W | −0.6V | - | 39 | 110 GHz | [11] |
| | | CVD | 1.55 | 136-395mA/W | −0.3V | - | ~230 | >40 GHz | [12] |
| | | | ~2 | 45~70 mA/W | | - | ~20 | >20 GHz | |
| | M-MoTe$_2$-M | mechanical exfoliation | 1.55 | 467 mA/W | −2V | 1-30 μW | 1.56×10$^7$ | 35 MHz | [13] |
| | M-PtSe$_2$-M | CVD | 1.55 | 12 mA/W | 8V | 0.05-0.8 mW (maybe beyond) | 3.5×10$^4$ | 35 GHz (Pulse) | [14] |
| | M-BP-M | mechanical exfoliation (Wet transfer) | 1.55 | 631 mA/W | 2 V | not given | 1.13×10$^3$ | 3 GHz | [15] |
| | M-Bi$_2$O$_2$Se-M | mechanical exfoliation | 1.26-1.31 | 3.5A/W | 2V | - | 3×10$^7$ | ~10 MHz | [16] |
| | graphene p-n homojunction | CVD (semidry transfer) | 1.55 | 6 V/W | 0V | not given | - | >67 GHz | [17] |
| | | mechanical exfoliation | 1.556 | ~ 90 V/W | 0V | 0.01-0.2 mW | - | 12 GHz | [18] |
| | Au-MoS$_2$ (Schottky PD) | mechanical exfoliation | 1.55 | 15.7mA/W | −0.3V | 0.03-0.23 mW | 1.05×10$^4$ | 1.37 GHz | [19] |
| 2DM-Heterostructure | MoTe$_2$-G[d] | mechanical exfoliation | 1.26-1.34 | 10-220 mA/W | −3~−0.6V | 0.02-0.25 mW (maybe beyond) | 10$^4$~10$^6$ | 12~46 GHz | [20] |
| | MoS$_2$/G-hBN-G | CVD | 1.55 | 0.24 A/W | 10 V | not given | 10$^7$ | 28 GHz (Pulse) | [21] |
| | Au-G-Si | CVD | 1.55 | 85 mA/W | −1V | 1.5-6 μW (maybe beyond) | 4.25×10$^6$ | - | [22] |
| | G-Si p-i-n heterojunction[e] | CVD | 1.55 | 3.4 mA/W (c.w. light)/ 11 mA/W (Pulse light) | 0V | 0.1-1 mW (Pulse light) | - | 29.4GHz (15 ps response time) />40GHz | [23] |
| | G-Si-G | CVD | 1.55 | 0.13-0.33 mA/W | 5-6 V | 3 μW-1.86 mW (maybe beyond) | 1.5×10$^6$ ~6.5×10$^6$ | 19-33 GHz | This work |
| | | | ~2 | 0.04-0.017 mA/W | | 0.17-6.81 mW (maybe beyond) | 3.92×10$^4$ ~3.79×10$^5$ | - | |

Note:

a) '-': the measured range responsivity changes. 'not given': no $P_{in}$ versus responsivity relation given.

b) '-': at zero bias voltage, the dark current is zero theoretically.

c) The bandwidths extracted from optical pulse measurement in time domain are annotated.

d) The highest performances are not obtained at the same device.

e) In this work, the responsivities are respectively 3.4 mA/W and 11 mA/W under c.w. light and pulse light injections. The bandwidth is 29.4GHz when estimated from the 15 ps response time, and is over 40 GHz according to the S$_{21}$ measurement.[23]

**Supplementary References:**